\documentclass[fleqn,usenatbib]{mnras}

\usepackage{newtxtext,newtxmath}

\usepackage[T1]{fontenc}

\DeclareRobustCommand{\VAN}[3]{#2}
\let\VANthebibliography\thebibliography
\def\thebibliography{\DeclareRobustCommand{\VAN}[3]{##3}\VANthebibliography}

\usepackage{ae,aecompl}
\usepackage{graphicx}
\usepackage{amsmath}
\usepackage{supertabular}
\usepackage{hyperref}
\usepackage{xcolor}
\usepackage{academicons}
\usepackage[normalem]{ulem}
\usepackage{multirow}

\def\mev{\;\text{MeV}}
\def\fm3{\;\text{fm}^{-3}}
\def\mev{\;\text{MeV}}
\newcommand{\Msun}{\,M_{\odot}}
\newcommand{\km}{\hbox{$\,{\rm km}$}}
\newcommand{\nicer}{{\it NICER}}

\title{On the Moment of Inertia of PSR J0737-3039 A from LIGO/Virgo and NICER}

\author[Miao et al.]{
Zhiqiang Miao$^{1}$\href{https://orcid.org/0000-0003-1197-3329}{\includegraphics[scale=0.07]{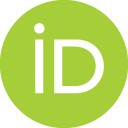}};
Ang Li$^{1}$\href{https://orcid.org/0000-0001-9849-3656}{\includegraphics[scale=0.07]{ORCIDiD_icon128x128.png}};
Zi-Gao Dai$^{2,3}$~\href{https://orcid.org/0000-0002-7835-8585}{\includegraphics[scale=0.07]{ORCIDiD_icon128x128.png}}\\
$^1$Department of Astronomy, Xiamen University, Xiamen, Fujian 361005, China; {\tt liang@xmu.edu.cn}\\
$^2$Department of Astronomy, School of Physical Sciences, University of Science and Technology of China, Hefei 230026, China\\
$^3$School of Astronomy and Space Science, Nanjing University, Nanjing 210023, China
}

\date{Accepted XXX. Received YYY; in original form ZZZ}

\begin{document}
\label{firstpage}
\pagerange{\pageref{firstpage}--\pageref{lastpage}}
\maketitle

\begin{abstract} 
We perform a Bayesian analysis of neutrons star moment of inertia by utilizing the available gravitational-wave data from LIGO/Virgo (GW170817 and GW190425) and mass-radius measurements from the Neutron Star Interior Composition Explorer (PSR J0030+0415 and PSR J0740+6620), incorporating the possible phase transition in the pulsar inner core. We find that the moment of inertia of pulsar A in the double pulsar binary J0737-3039 is $\sim1.30\times10^{45}\,{\rm g\,cm^2}$, which only slightly depends on the employed hadronic equation of states. We also demonstrate how a moment of inertia measurement would improve our knowledge of the equation of state and the mass-radius relation for neutron stars and discuss whether a quark deconfinement phase transition is supported by the available data and forthcoming data that could be consistent with this hypothesis. 
We find that if pulsar A is a quark star, that its moment of inertia is a large value of $\sim1.55\times10^{45}\,{\rm g\,cm^2}$ suggesting the possibility of distinguishing it from (hybrid-)neutron stars with measurements of PSR J0737-3039A moment of inertia. 
We finally demonstrate the moment-of-inertia–compactness universal relations and provide analytical fits for both (hybrid-)neutron star and quark star results based on our analysis.
\end{abstract}


\section{Introduction}           
\label{sect:intro}

Neutron stars are the densest and smallest stars observed in the Universe. Observations of neutron stars provide valuable information on the internal structure of these objects and, therefore, on the nuclear equation of state (EOS) of dense matter in their interior.
Various observables are tightly related with microphysics inside the star and can help to set reliable constraints on its EOS. Among them, there are the mass~\citep{2010Natur.467.1081D,2013Sci...340..448A,2020NatAs...4...72C,2021ApJ...915L..12F}, the radius~\citep{2019ApJ...887L..21R,2021ApJ...918L..27R,2019ApJ...887L..24M,2021ApJ...918L..28M}, and the tidal deformability~\citep{2017PhRvL.119p1101A,2018PhRvL.121p1101A,2020ApJ...892L...3A}. 
In addition, measurements are planned for the moment of inertia (MOI) of PSR J0737-3039 A (hereafter A)~\citep{2004Sci...303.1153L}, the $1.338\Msun$ primary component of the first double pulsar system PSR J0737-3039~\citep{2003Natur.426..531B}, based on the long-term pulsar timing to determine the periastron advance due to relativistic spin-orbit coupling~\citep{1988NCimB.101..127D}.
The MOI of A is expected to be measured with $\sim 10\%$ accuracy within the next decade, 
which will provide complementary constraints on the EOS~\citep{2005MNRAS.364..635B,2005ApJ...629..979L,2020ApJ...901..155G,2020MNRAS.497.3118H}. 

The neutron star MOI has been theoretically calculated using a wide range of EOS models~\citep[e.g.,][]{1999ApJ...525..950Y,2004ApJ...617L.135M,2005ApJ...629..979L,2005MNRAS.364..635B,2007NuPhA.792..341R,2007PhLB..654..170K,2009PhRvD..79l4032R,2013MNRAS.433.1903U,2015ChPhL..32g9701L,2016ApJS..223...16L,2019ApJ...878..159M,2019AIPC.2127b0020S,2019JPhG...46c4001W,2020EPJA...56...32L,2020PhRvD.101d3023M}.
Several studies have been performed using EOS models, including those incorporating non-nucleon baryonic components (such as $\Delta$-isobars, hyperons and antikaon condensates, etc.) 
~\citep[e.g.,][]{2007PhLB..654..170K,2018EPJA...54...26B,2019JPhG...46a5202S,2019ApJ...883..168R,2020PhRvD.101c4017F,2020MNRAS.499..914R}.

In several recent studies, the MOI of A has been predicted based on EOSs constrained using physical stability arguments~\citep{2016PhRvC..93c2801R}, neutron star observations
~\citep{2015PhRvC..91a5804S,2018ApJ...868L..22L,2019PhRvD..99l3026K,2020ApJ...892...55J,2021PhRvL.126r1101S}, or the results of nuclear experiments~\citep{2008ApJ...685..390W,2019PhRvC.100c5802L}.
However, a phase transition was not explicitly taken into account in these previous studies.
In the inner stellar core, there may be a possibility that the hadron pressure is so high that nucleons should reasonably melt down to the deconfined phase, and EOSs with an early phase transition at $\sim2n_0$ (where $n_0=0.16\,{\rm~fm^{-3}}$ denotes the nuclear saturation density) are compatible with current neutron star observations~\citep{2021PhRvL.126f1101A,2020JHEAp..28...19L,2021ApJ...913...27L,2020ApJ...904..103M,2021PhRvC.103c5802X}, therefore a description of the high-density core matter in terms of only nucleons might be inadequate for the the MOI study of A. 

Furthermore, in the extreme case of quark deconfinement phase transition, neutron stars may convert completely to quark stars~\citep[e.g.,][]{1987PhLB..192...71O,1994PhRvD..49.2698O,2002A&A...390L..39O,2011PhRvD..84h3002H}, resulting in various astrophysical implications, such as neutrino and gravitational waves emission~\citep[e.g.,][]{2013PhRvD..87j3007P,2016EPJA...52...58B} as well as short gamma-ray bursts~\citep[e.g.,][]{1996PhRvL..77.1210C,2002A&A...390L..39O,2003ApJ...586.1250B}. 
There are also discussions on the trigger of such transition due to self-annihilation of dark matter~\citep{2010PhRvL.105n1101P} and the characteristic delay due to the transition that could lead to observational signatures helping distinguish different central engine models of short gamma-ray bursts~\citep{2013ApJ...768..145P}. Therefore, it is interesting and meaningful to consider the possibility of A as a quark star and discuss all three different MOI results when assuming A is a pure neutron star, a hybrid star, and a pure quark star. By doing so, we are in the right place to investigate whether a future MOI measurement to $10\%$ precision is sufficient to distinguish among the three qualitatively different models. 
What particularly interesting is that, since the quark matter in hybrid stars and quark stars are described under different stability conditions, and both of them can't coexist stably~\citep{2005PrPNP..54..193W}, a confirmation (or exclusion) of either of them can shed light on the remaining-unsolved QCD at finite baryon density~\citep{2021PhRvL.127p2003G}. 

This paper is organized as follows.
In Section~\ref{sec:eos}, we introduce the EOSs used to model the hadronic phase of neutron stars and a general parameterization for describing the phase transition and the quark matter phase of neutron stars.
The neutron star observations used and the Bayesian analysis method are detailed in Section~\ref{sec:analysis}.
We present our results and a discussion in Section~\ref{sec:res} and summarize our paper in Section~\ref{sec:summary}.

\section{Neutron stars with or without a quark core}
\label{sec:eos}

Following our previous study~\citep{2021ApJ...913...27L}, we describe neutron stars with and without a quark core below.  

For neutron stars without a quark core, we employ the unified QMF~\citep{2018ApJ...862...98Z} and DD2~\citep{2016PhRvC..94c5804F} EOS models for soft and stiff hadronic matter, while using the standard BPS EOS for the neutron star outer crust~\citep{1971ApJ...170..299B}.
Solving the Tolman-Oppenheimer-Volkoff (TOV) equations yields the neutron star maximum mass as $M_{\rm TOV}^{\rm QMF}=2.07\Msun$ and $M_{\rm TOV}^{\rm DD2}=2.42\Msun$, with the radius of a typical $1.4\Msun$ star as $R_{1.4}^{\rm QMF}=11.77\,{\rm km}$ and $R_{1.4}^{\rm DD2}=13.17\,{\rm km}$, respectively.
Furthermore, as the rotation frequency of A, $\Omega_{\rm A} = 276.8\,{\rm Hz}$~\citep{2003Natur.426..531B} is far smaller than the Kepler frequency, 
one can employ the slow rotation approximation~\citep{1967ApJ...150.1005H} to calculate the MOI of the $1.338$-solar-mass A, which is $I_A^{\rm QMF}=1.33\times10^{45}\,{\rm g\,cm^2}$ and $I_A^{\rm DD2}=1.60\times10^{45}\,{\rm g\,cm^2}$ using the two EOSs.

For neutron stars with a quark core, we implement a sharp first-order phase transition between the hadronic and quark phases by employing the ``constant-speed-of-sound'' (CSS) parameterization ~\citep{2013PhRvD..88h3013A}, while maintaining the hadronic EOS as QMF or DD2.
The full EOS can be expressed as follows (we use units where $\hbar=c=1$):
\begin{equation}
\varepsilon(p) = \left\{\!
\begin{aligned}
&\varepsilon_{\rm HM}(p)\ , & p<p_{\rm trans} \\ 
&\varepsilon_{\rm HM}(p_{\rm trans})+\Delta\varepsilon+c_{\rm QM}^{-2} (p-p_{\rm trans})\ , & p>p_{\rm trans} 
\end{aligned}
\right.
\end{equation}
where $p_{\rm trans}$ is the pressure of the transition, $\Delta\varepsilon$ is the discontinuity in the energy density, and $c_{\rm QM}$ is the sound speed in quark matter. 
Correspondingly, three dimensionless parameters are chosen as the EOS parameters of hybrid stars, namely, the transition density $n_{\rm trans}/n_0$, the transition strength $\Delta\varepsilon/\varepsilon_{\rm trans}$, and the sound speed squared in quark matter $c_{\rm QM}^2$, where $n_{\rm trans}\equiv n_{\rm HM}(p_{\rm trans})$ and $\varepsilon_{\rm trans}\equiv \varepsilon_{\rm HM}(p_{\rm trans})$.

In the following, within the context of a first-order phase transition from hadronic matter to quark matter in the neutron star interior, we intend to make a direct estimate of the MOI of A based on a Bayesian analysis of observational data for the tidal deformability of the GW170817~\citep{2017PhRvL.119p1101A} and GW190425~\citep{2020ApJ...892L...3A} binary neutron star mergers, as detected by LIGO/Virgo, and the masses and radii of PSR J0030+0451~\citep{2000ApJ...545.1007L,2000ApJ...545.1015B} and PSR J0740+6620~\citep{2014ApJ...791...67S}, as detected by \nicer~\citep{2019ApJ...887L..24M,2019ApJ...887L..21R,2021ApJ...918L..28M,2021ApJ...918L..27R}.
The EOS parameters are varied to obtain a 3D parameter space for the EOSs, and a Bayesian analysis is performed to select preferred EOSs based on the neutron star observations from LIGO/Virgo and \nicer. 
The corresponding study on quark stars are then followed (see below in Sec.~\ref{sec:q}). 
The Bayesian analysis framework is introduced in the next section.

\section{Observational constraints and Bayesian analysis}
\label{sec:analysis}

In the present study, our observational data set $\boldsymbol{d}$ contains two gravitational wave events (GW170817 and GW190425) and two \nicer~ mass-radius measurements of pulsars (PSR J0030+0451 and PSR J0740+6620). 
Assuming each dataset in $\boldsymbol d$ is independent of one another, the total likelihood can be expressed as:
\begin{equation}\label{eq:posterior}
\begin{split}
    \mathcal{L}(\boldsymbol{d}|\boldsymbol\theta,\boldsymbol p_c)= \prod_i \mathcal{L}_{{\rm GW},i}\times \prod_j \mathcal{L}_{{\rm NICER},j}\ ,
\end{split}
\end{equation}
where the priors of the parameters are $p(\boldsymbol\theta,\boldsymbol p_c)=p(\boldsymbol\theta)p(\boldsymbol p_c)$, with $\boldsymbol p_c$ being the central pressure of the \nicer's pulsars. We employ the same priors as in our previous study~\citep{2021ApJ...913...27L} for EOS parameters $\boldsymbol\theta = \{n_{\rm trans}/n_0,\Delta\varepsilon/\varepsilon_{\rm trans}, c_{\rm QM}^2\}$. 

{\bf GW170817 and GW190425.}
We calculate the likelihood of each gravitational event through a high-precision interpolation of the likelihood developed in~\citet{2020MNRAS.499.5972H} from fitting the strain data released by LIGO/Virgo, which is encapsulated in the python package \textsf{toast}, i.e.,
\begin{equation}
     \mathcal{L}_{{\rm GW},i}= F_i(\Lambda_1, \Lambda_2, \mathcal M_c, q)\ ,
\end{equation}
where $F_i(\cdot)$ is the interpolation function of the gravitational wave event $i$. The chirp mass $\mathcal M_c \equiv (M_1M_2)^{3/5}/(M_1+M_2)^{1/5}$ and the mass ratio $q\equiv M_2/M_1$. The tidal deformabilities can be derived from the component masses, i.e., $\Lambda_1 = \Lambda_1(\boldsymbol\theta; M_1)$ and $\Lambda_2 = \Lambda_2(\boldsymbol\theta; M_2)$.
We mention here that since the likelihood involves no EOS assumption about the phase state of the merging neutron stars, we can adopt it for the study of both (hybrid) neutron stars and quark stars.

{\bf \nicer~mass-radius measurements.}
Recently, the \nicer~collaboration reported simultaneous mass-radius measurements of PSR J0030+0451 and PSR J0740+6620. The results for PSR J0030+0451 were $M=1.34_{-0.16}^{+0.15}\Msun$, $R=12.71_{-1.19}^{+1.14}\,{\rm km}$~\citep{2019ApJ...887L..21R} or $M=1.44_{-0.14}^{+0.15}\Msun$, $R=13.02_{-1.06}^{+1.24}\,{\rm km}$~\citep{2019ApJ...887L..24M}. The results for PSR J0740+6620 were $M=2.072_{-0.066}^{+0.067}\Msun$, $R=12.39_{-0.98}^{+1.30}\,{\rm km}$~\citep{2021ApJ...918L..27R} or $M=2.062_{-0.091}^{+0.090}\Msun$, $R=13.71_{-1.50}^{+2.61}\,{\rm km}$~\citep{2021ApJ...918L..28M}. To incorporate each of these measurements, we write the likelihood as:
\begin{equation}\label{eq:llh_NICER}
    \mathcal{L}_{{\rm NICER},j} = {\rm KDE}(M,R|\boldsymbol S_j)\ ,
\end{equation}
where the right-hand side is a kernel density estimation of the posterior samples $\boldsymbol S_j$ of the mass-radius measurement $j$. Similar to~\citet{2021ApJ...913...27L}, we use the ST+PST model samples from~\citet{riley_thomas_e_2019_3386449} for PSR J0030+0451. For PSR J0740+6620, we use the \nicer~x XMM samples from~\citet{riley_thomas_e_2021_4697625}. The mass and radius $(M, R)$ are mapped from the EOS parameters $\boldsymbol\theta$ and the central pressure $\boldsymbol p_c$ through the TOV equations.
For PSR J0030+0451, we set $p_c$ (in unit ${\rm dyn/cm^2}$) to uniformly distribute in logarithmic space between $\log(p_c)=34.5$ and $\log(p_c)=35.5$, while for PSR J0740+6620, $p_c$ uniformly distributes in logarithmic space from $\log(p_c)=35$ to $\log(p_c)=36$. In addition, we consider parameter sets $\{\boldsymbol\theta,\boldsymbol p_c\}$ that correspond to stable neutron stars, i.e., those $\{\boldsymbol\theta,\boldsymbol p_c\}$ correspond to unstable neutron stars should be rejected in sampling.
See more discussions in our previous study~\citep{2021ApJ...913...27L}.

The posterior distribution is proportional to the product of the prior and the likelihood function, i.e.,
\begin{equation}
    p(\boldsymbol\theta, \boldsymbol p_c|\boldsymbol d) \propto p(\boldsymbol\theta,\boldsymbol p_c)\mathcal{L}(\boldsymbol{d}|\boldsymbol\theta,\boldsymbol p_c)\ .
\end{equation}
In practice, we sample from the posterior distribution by using the nested sampling software \textsf{PyMultiNest}~\citep{2014A&A...564A.125B}. We finally obtain a total of approximately 3,000 posterior samples of the CSS model parameters for both QMF+CSS and DD2+CSS. These EOS parameter samples are used in the study of (hybrid)neutron star properties.

\section{Results and discussion}
\label{sec:res}

\begin{figure}
        \includegraphics[width=3.3in]{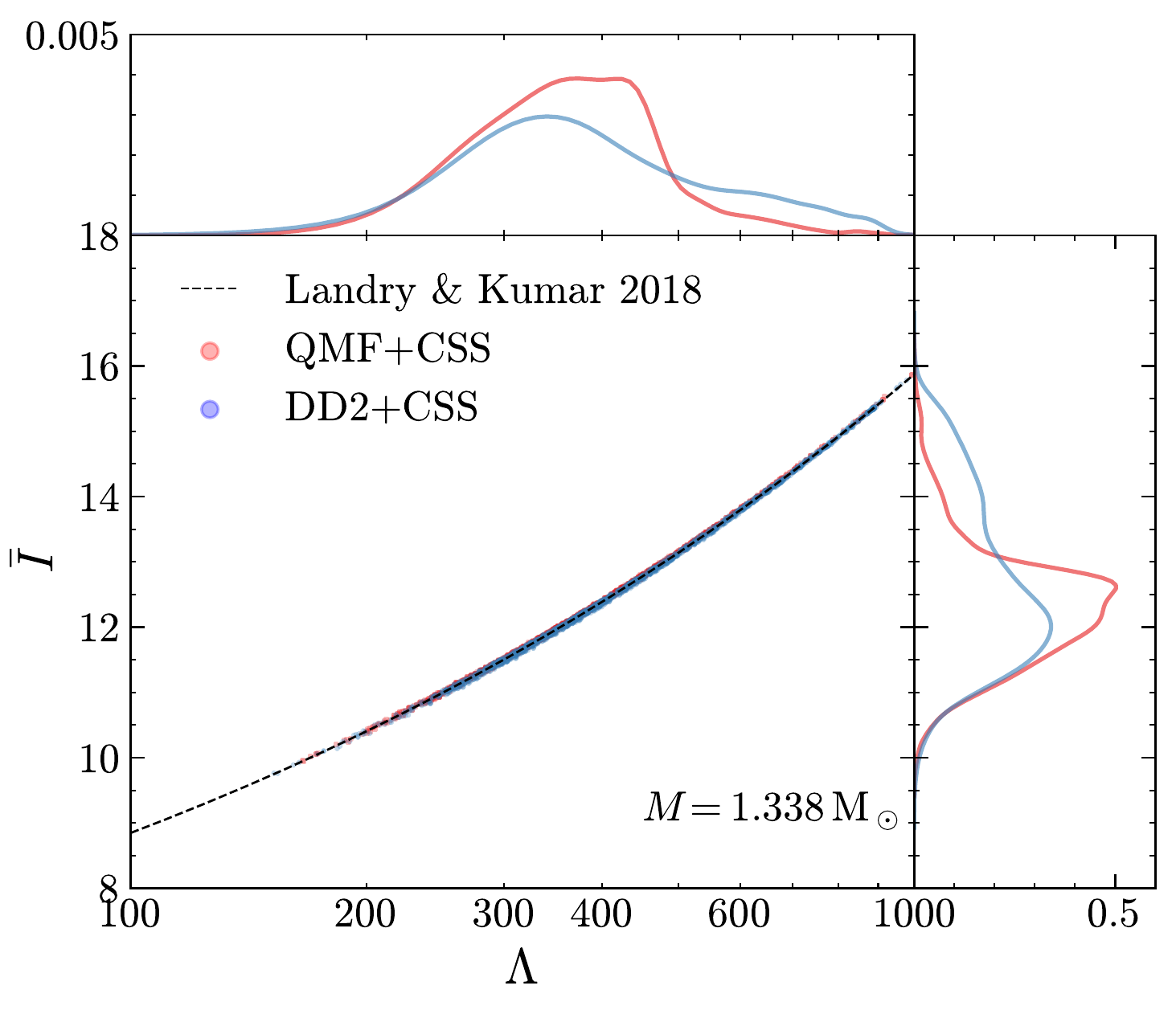}
         \vskip-2mm
    \caption{Posterior distribution of $\bar I \equiv I/M^3$ vs. $\Lambda$ for a $1.338\Msun$ star. The dots are calculated from the corresponding EOS parameter posterior samples. The marginalized posterior distributions of $\Lambda$ and $\bar I$ are shown in the upper and right panel, respectively. The black dashed line shows the fitting formula from~\citet{2018ApJ...868L..22L}, updated from the original relation of \citet{2013PhRvD..88b3009Y} with a larger set of EOSs.  
    }
         \vspace{-0.5cm}
    \label{fig:I-Love}
\end{figure}

\subsection{MOI of PSR J0737-3039 A}
\label{sec:MOI prediction}

Fig.~\ref{fig:I-Love} presents the posterior distributions of the dimensionless MOI $\bar{I}\equiv I/M^3$ vs. the tidal deformability $\Lambda$. 
The analysis is performed for a fixed star mass $M=1.338\Msun$, corresponding to A.
Our results can be effectively reconciled with the universal relation of neutron stars (shown as a black line in Fig.~\ref{fig:I-Love})~\citep{2013PhRvD..88b3009Y,2018ApJ...868L..22L}, with an error less than 0.5\%, indicating that the I-Love relation remains valid within the present hybrid star scenario~\citep[see also in, e.g.,][]{2020ApJ...904..103M}.
Such universality is expected since hybrid stars and neutron stars differ only in the high-density core part and it is the low-density crust part that mainly determines the $\bar{I}$-$\Lambda$ curves with different EOSs~\citep{2013PhRvD..88b3009Y}.

\begin{figure}
        \includegraphics[width=3.3in]{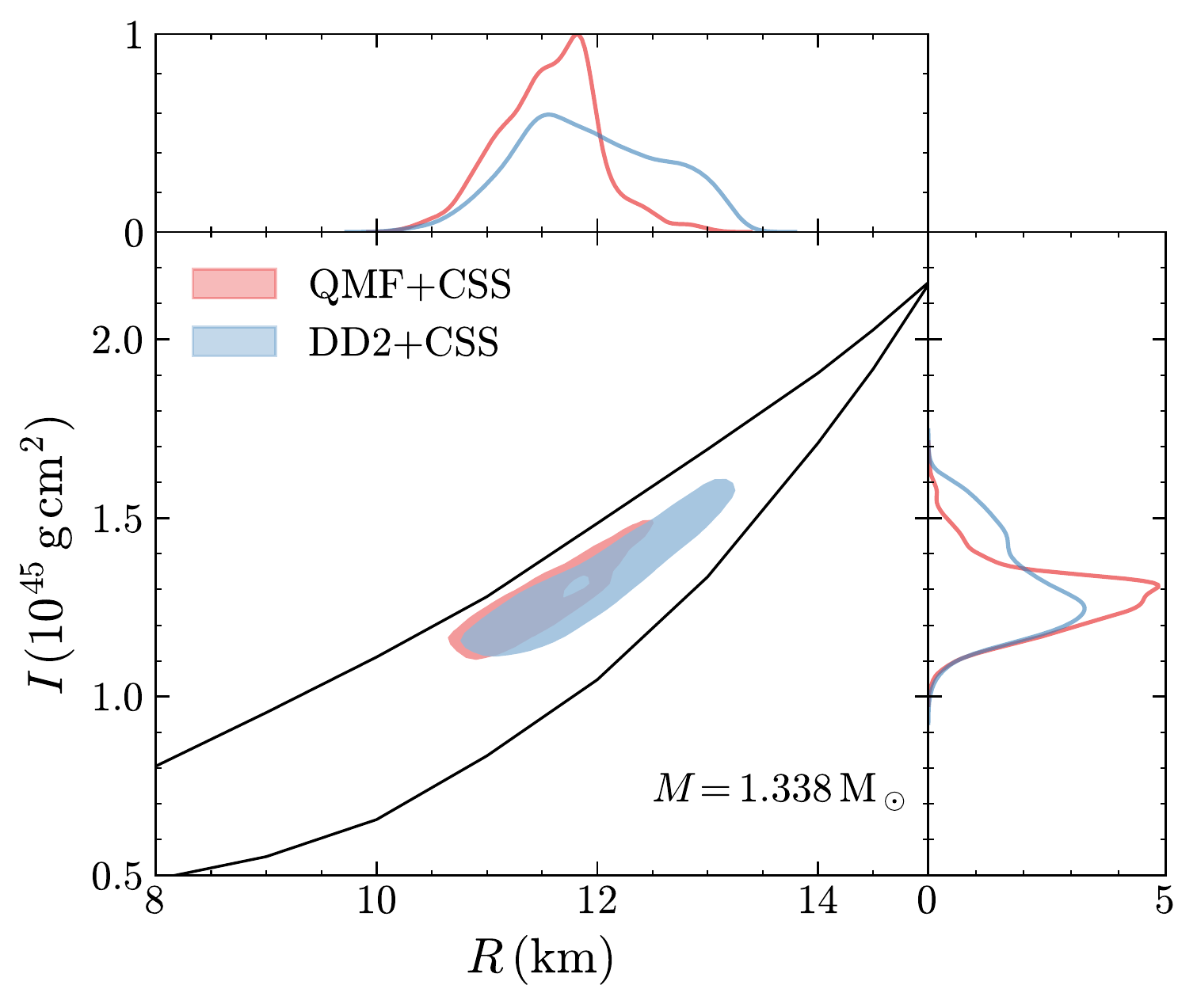}
         \vskip-2mm
    \caption{Theoretical bounds on the MOI of J0737-3039 A as a function of its radius, within both QMF+CSS and DD2+CSS (90\% credible interval). The marginalized posterior distributions of $R$ and $I$ are shown in the upper and right panel, respectively. The black curve corresponds to an extreme bound from \citet{2016PhRvC..93c2801R}, assuming the AP4 EOS up to nuclear saturation density $n_0$ and a constant-density core inside the neutron stars. 
    }
    \vspace{-0.5cm}
    \label{fig:R-I}
\end{figure}

\begin{table}
  \centering
  \caption {Theoretical predictions on the MOI of J0737-3039 A ($M=1.338\Msun$), given in $10^{45}\,{\rm g\,cm^2}$. Except the values of~\citet{2008ApJ...685..390W}, the quoted error bars refer to symmetric 90\% credible intervals about the median. We mention here that the current $90\%$ upper limit from recent double pulsar timing measurements is $3\times10^{45}\,\rm g\,cm^2$~\citep{2021PhRvX..11d1050K}. 
  }
    \setlength{\tabcolsep}{0.8pt}
\renewcommand\arraystretch{1.3}
\begin{tabular*}{\hsize}{@{}@{\extracolsep{\fill}}lr@{}}
    \hline
    Reference & $I_A$\\
    \hline
    \citet{2008ApJ...685..390W} & $[1.30, 1.63]$   \\ 
    \citet{2018ApJ...868L..22L} & $1.15_{-0.24}^{+0.38}$  \\    
    \citet{2019PhRvC.100c5802L} & $1.36_{-0.26}^{+0.12}$ \\
    \citet{2020ApJ...892...55J} & $1.35_{-0.14}^{+0.26}$ \\
    \citet{2021PhRvL.126r1101S} &$1.68_{-0.48}^{+0.53}$  \\
    This work (QMF+CSS) & $1.27_{-0.14}^{+0.18}$  \\
    This work (DD2+CSS) & $1.29_{-0.15}^{+0.26}$  \\    
    \hline
  \end{tabular*}
          \vspace{-0.2cm}
  \label{tb:MOIcomparison}
\end{table}

In Fig.~\ref{fig:R-I}, we show the 90\% bounds on the MOI for A as a function of the A radius.
Also shown is the extreme bound from~\citet{2016PhRvC..93c2801R}, which is derived by placing one or two constant-density cores inside the neutron stars based on stability arguments, whereas the outer layer (with a density lower than the nuclear saturation density) is described by the AP4 EOS.
The EOS modelling results based on currently available neutron star observations fall within a fairly narrow region of the $I-R$ plane.

In Table~\ref{tb:MOIcomparison}, we compare our MOI prediction for A with those of previous studies.
Using EOS-insensitive relations,~\citet{2018ApJ...868L..22L}  resulted in an MOI of $\sim1.15\times10^{45}\,{\rm g\,cm^2}$ based on the GW170817 tidal measurement. In~\citet{2021PhRvL.126r1101S}, a larger MOI of $\sim1.68\times10^{45}\,{\rm g\,cm^2}$ was obtained based on the \nicer~ mass-radius measurement of PSR J0030+0451 at 90\% credible interval.
In \citet{2020ApJ...892...55J}, both data sets were used with parameterized EOSs to obtain an MOI between $\sim 1.35\times10^{45}\,{\rm g\,cm^2}$.
In the present study, we not only explicitly include the possibility of a phase transition in the EOS prior but also incorporate two additional sources, GW190425 and PSR J0740+6620, into the analysis.
Our results indicate that $I_A=1.27_{-0.14}^{+0.18}\times10^{45}\,{\rm g\,cm^2}$ (or $I/M_{\rm A}^{3/2} = 41.3_{-4.5}^{+5.8}{\rm km^2}/\Msun^{1/2}$) within QMF+CSS or $1.29_{-0.15}^{+0.26}\times10^{45}\,{\rm g\,cm^2}$ (or $I/M_{\rm A}^{3/2} = 41.9_{-4.9}^{+8.4}{\rm km^2}/\Msun^{1/2}$) within DD2+CSS at $90\%$ credible interval.
Predictions have also been made based on EOSs constrained by relativistic heavy-ion collisions~\citep{2008ApJ...685..390W} or microscopic EOSs consistent with empirical data for finite nuclei~\citep{2019PhRvC.100c5802L}, as shown in Table~\ref{tb:MOIcomparison}.

\begin{figure}
        \includegraphics[width=3.3in]{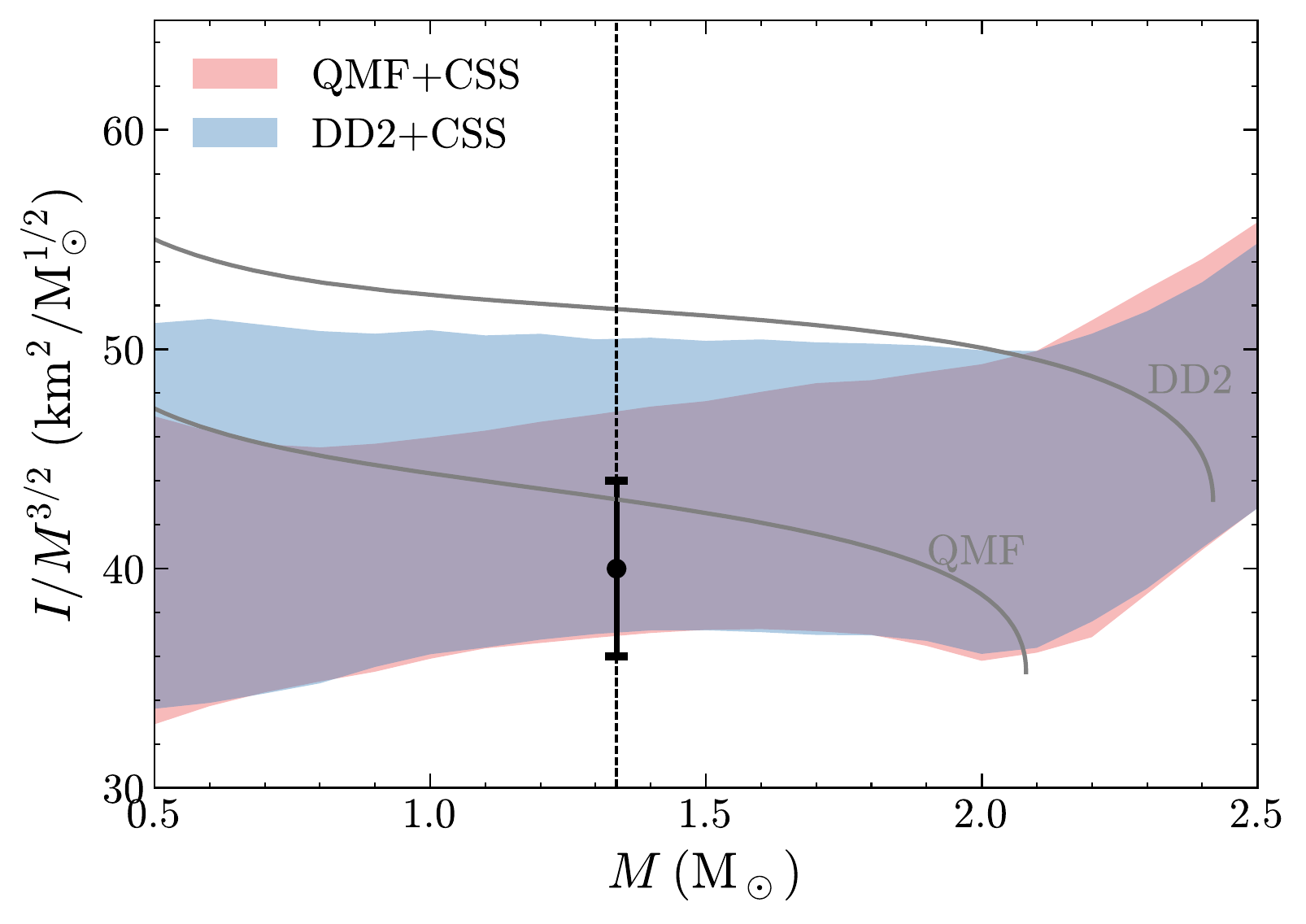}
         \vskip-2mm
    \caption{The $90\%$ credible regions of $I/M^{3/2}$ as a function of $M$, transformed from the EOS parameter posteriors. Also shown in solid grey lines are the results of two neutron star EOSs (QMF and DD2). The vertical line corresponds to $M=1.338\Msun$, while the error bar represents a hypothetical $10\%$ precision measurement centered on $I/M_{\rm A}^{3/2}=40\,{\rm km^2}/\Msun^{1/2}$.
    }
        \vspace{-0.5cm}
    \label{fig:M-I}
\end{figure}

For subsequent calculations, Fig~\ref{fig:M-I} shows the posterior distributions of $I/M^{3/2}$ as a function of the mass $M$.
The scaled MOI is found to lie within the range $I/M^{3/2}\sim35-50\,{\rm km^2}/\Msun^{1/2}$.
Note that in \citet{2015PhRvC..91a5804S}, a similar and somewhat narrower range of $\sim 35-45\,{\rm km^2}/\Msun^{1/2}$ was calculated based on two sets of parameterized EOSs (which were denoted as Models A and C) obtained from analyzing several mass-radius measurements from photospheric radius expansion (PRE) bursts or quiescent low-mass X-ray binaries.

Below, we use the implications of the LIGO/Virgo and \nicer~ data for the MOI of A (as shown in Fig.~\ref{fig:M-I}) to analyze the available MOI data and the mocked data consisting of a future MOI measurement of A with an accuracy of $10\%$.
In Sec.~\ref{sec:oldI}, we demonstrate how the available MOI measurements add no new information on the EOS of neutron star matter due to its huge uncertainties. 
In Sec.~\ref{sec:artI}, we present updated constraints on the EOS and the corresponding mass-radius relations of the stars from a mocked MOI data with an accuracy of $10\%$.
In Sec.~\ref{sec:withq}, we explore further whether a phase with deconfined quarks is supported by current and future observational data.
In Sec.~\ref{sec:q}, we further discuss the possibility of distinguishing quark stars from (hybrid-)neutron stars with the MOI measurements of A.

\subsection{Constraints with available MOI measurements} \label{sec:oldI}

\begin{figure}
    \includegraphics[width=3.3in]{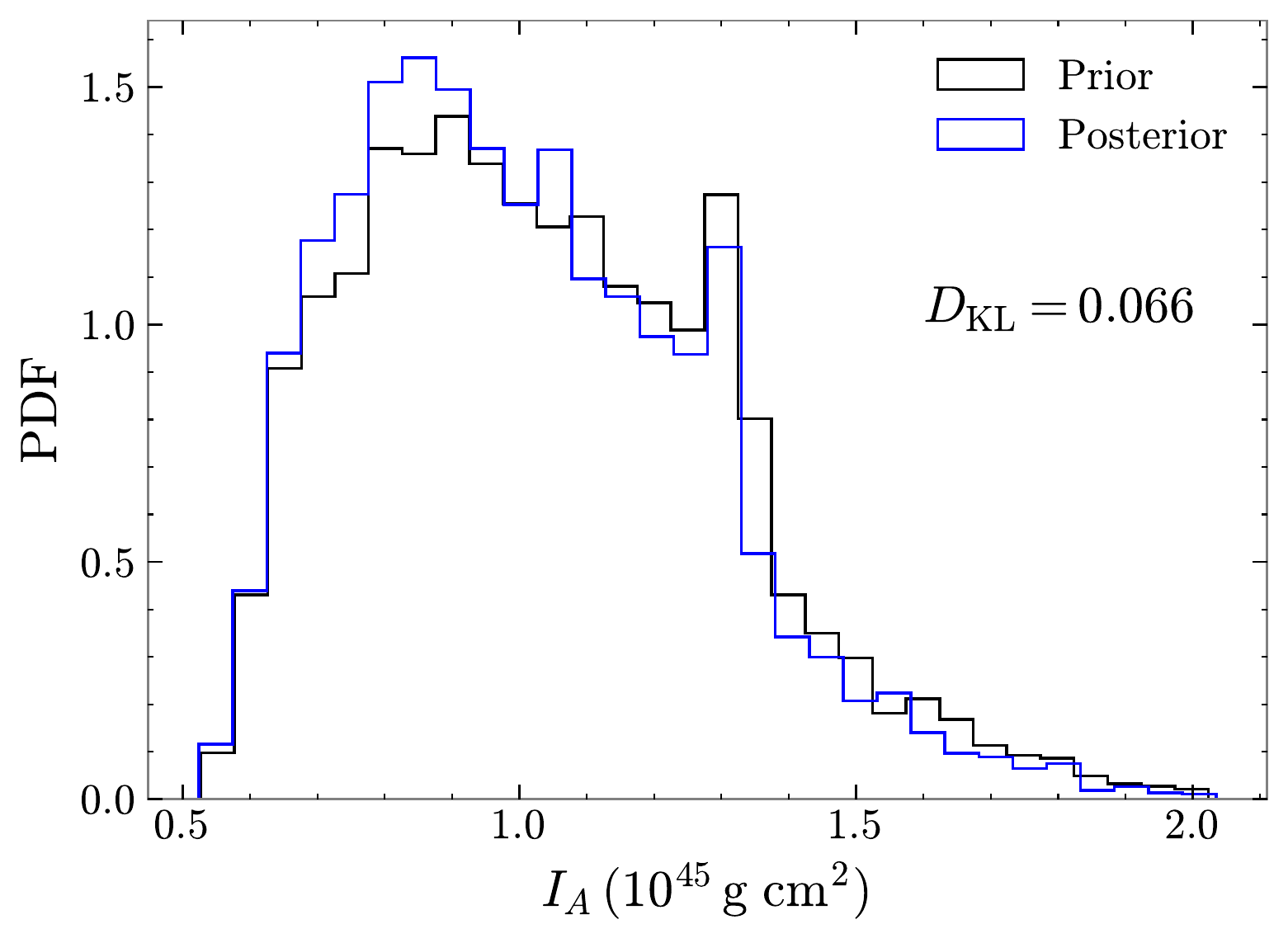}
         \vskip-2mm
    \caption{Prior and posterior distributions of the MOI of pulsar A. The prior of $I_A$ is obtained from the QMF+CSS model with the prior of the CSS model parameters in Sec.~\ref{sec:analysis}, while the posterior is conditioned on the the MOI result from~\citet{2021PhRvX..11d1050K}. The Kullback–Leibler divergence, $D_{\rm KL}$, from prior to posterior in bits for $I_A$ is also shown. 
    }
        \vspace{-0.5cm}
    \label{fig:PDF_IA}
\end{figure}

Recently, an MOI measurement of pulsar A, with a 90\% upper limit $I_A<3\times10^{45}\,{\rm g\ cm^2}$ is reported in \citet{2021PhRvX..11d1050K}, by fitting the post-Keplerian parameters in the binary system.
It is mentioned in~\citet{2021PhRvX..11d1050K} that such a measurement 
gives a loose upper limit for A’ radius of 22 km (with 90\% confidence) and is regarded as not constraining to the EOS, in comparison with the accurate observations from, e.g.,  LIGO/Virgo, where $R_{\rm 1.4} \leqslant 13.5$ km is obtained~\citep{2018PhRvL.121p1101A}.  
We here report a similar conclusion in Fig.~\ref{fig:PDF_IA}, where we compare the prior distribution and the posterior distribution of the MOI of pulsar A. 

The prior of $I_A$ is calculated from the QMF+CSS model with the prior of the CSS model parameters in Sec.~\ref{sec:analysis}. 
The posterior is obtained from the prior and the likelihood of the MOI measurement reported in~\citet{2021PhRvX..11d1050K}. In practice, we equal the likelihood function to the posterior distribution of the MOI of A reported in~\citet{2021PhRvX..11d1050K}, which can be well fitted by a Gaussian distribution with mean $\mu_I=0.283\times 10^{45}\,{\rm g\ cm^2}$ and standard deviation $\sigma_I=1.705\times 10^{45}\,{\rm g\ cm^2}$, i.e.,
\begin{equation}
    \mathcal{L}_I = \frac{1}{\sqrt{2\pi}\sigma_I}\exp\left\{-\frac{(I_A-\mu_I)^2}{2\sigma_I^2}\right\}\ .
\end{equation}
Again, we sample from posterior distribution and the result is shown in Fig.~\ref{fig:PDF_IA}.
We also calculate the Kullback–Leibler divergence from prior to posterior in bits for $I_A$, $D_{\rm KL}=0.066$.
The small value of $D_{\rm KL}$ and the visual comparison of the prior and posterior indicate that the available MOI measurements do provide no helpful constraints on the EOS, with respect to the LIGO/Virgo and \nicer~data. 

\begin{table*}
  \centering
  \caption {The resulting sound speed squared in quark matter $c_{\rm QM}^2$, the transition pressure $p_{\rm trans}$ and discontinuity on the energy density $\Delta\varepsilon$, the maximum mass $M_{\rm TOV}$ and the radius of a $1.4\Msun$ star $R_{1.4}$ are collected for both cases of QMF+CSS and DD2+CSS, from our previous work~\citep{2021ApJ...913...27L} (in the 2nd column) and the present work (in the 3rd column). 
  In the last four columns, we show the results of four tests when a mocked MOI measurement is added, being $I/M_{\rm A}^{3/2}=35, 40, 45, 50\,{\rm km^2}/\Msun^{1/2}$, respectively. The quoted error bars refer to symmetric 90\% credible intervals about the median. 
  The Bayes factors of the hybrid star against the normal neutron star, $\mathcal{B}_{\rm NS}^{\rm HS}$, are also shown for each case. See text for details.}
    \setlength{\tabcolsep}{0.8pt}
\renewcommand\arraystretch{1.3}
\begin{tabular*}{\hsize}{@{}@{\extracolsep{\fill}}ccccccc@{}}
    \hline
   \multicolumn{1}{c}{\multirow{2}*{}} & GW170817+ PSR J0030
   & GW170817+GW190425 & \multicolumn{4}{c}{+Mocked MOI measurement}\\ 
   \cline{4-7}
   &(Li et al. 2021) & +PSR J0030+ PSR J0740 & Test i & Test ii & Test iii & Test iv\\
    \hline 
    \multicolumn{7}{c}{QMF+CSS}\\
    \hline
    $c_{\rm QM}^2$ 
    & $0.81_{-0.28}^{+0.17}$ & $0.73_{-0.30}^{+0.24}$
    & $0.77_{-0.27}^{+0.21}$ & $0.74_{-0.27}^{+0.24}$
    & $0.71_{-0.28}^{+0.26}$ & $0.70_{-0.30}^{+0.27}$\\ 
    $p_{\rm trans}/{\rm MeV\,fm^{-3}}$
    & $17.6_{-14.0}^{+37.9}$ & $16.8_{-13.9}^{+42.5}$
    & $17.1_{-14.1}^{+31.2}$ & $19.3_{-16.4}^{+38.1}$
    & $17.9_{-15.0}^{+41.9}$ & $12.8_{-10.3}^{+45.8}$\\ 
    $\Delta\varepsilon/{\rm MeV\,fm^{-3}}$
    & $87.0_{-72.8}^{+84.2}$ & $94.9_{-84.0}^{+87.8}$
    & $134.5_{-78.4}^{+72.8}$ & $101.9_{-82.4}^{+81.8}$
    & $78.5_{-68.9}^{+86.4}$ & $63.6_{-55.8}^{+76.0}$\\ 
    $M_{\rm TOV}/\Msun$ 
    & $2.36_{-0.26}^{+0.49}$ & $2.23_{-0.21}^{+0.42}$
    & $2.22_{-0.20}^{+0.29}$ & $2.24_{-0.22}^{+0.35}$
    & $2.23_{-0.22}^{+0.44}$ & $2.25_{-0.24}^{+0.54}$\\
    
     $R_{1.4}/{\rm km}$     
    & $11.70_{-0.74}^{+0.85}$ & $11.61_{-0.76}^{+0.75}$
    & $11.21_{-0.60}^{+0.60}$ & $11.52_{-0.67}^{+0.48}$
    & $11.74_{-0.68}^{+0.60}$ & $11.88_{-0.61}^{+0.75}$\\ 

     $\mathcal{B}_{\rm NS}^{\rm HS}$     
    &1.49 &0.63 &1.77  &0.66 & 0.43 & 0.44 \\ 
    \hline
    \multicolumn{7}{c}{DD2+CSS}\\
    \hline
    $c_{\rm QM}^2$ 
    & $0.80_{-0.29}^{+0.18}$ & $0.71_{-0.29}^{+0.26}$
    & $0.78_{-0.26}^{+0.20}$ & $0.73_{-0.27}^{+0.25}$
    & $0.69_{-0.28}^{+0.27}$ & $0.68_{-0.28}^{+0.29}$\\ 
    $p_{\rm trans}/{\rm MeV\,fm^{-3}}$
    & $14.1_{-9.6}^{+18.2}$ & $15.2_{-11.3}^{+21.1}$
    & $12.8_{-8.9}^{+15.6}$ & $12.9_{-9.2}^{+18.9}$
    & $15.2_{-11.4}^{+20.5}$ & $20.8_{-16.7}^{+18.0}$\\ 
    $\Delta\varepsilon/{\rm MeV\,fm^{-3}}$
    & $141.2_{-89.4}^{+91.4}$ & $149.9_{-96.6}^{+83.3}$
    & $195.7_{-53.0}^{+68.1}$ & $168.0_{-70.9}^{+67.8}$
    & $134.0_{-74.8}^{+77.9}$ & $106.1_{-73.3}^{+83.1}$\\ 
    $M_{\rm TOV}/\Msun$ 
    & $2.39_{-0.28}^{+0.42}$ & $2.23_{-0.21}^{+0.39}$
    & $2.23_{-0.21}^{+0.27}$ & $2.25_{-0.22}^{+0.35}$
    & $2.25_{-0.22}^{+0.41}$ & $2.22_{-0.20}^{+0.45}$  \\ 
     $R_{1.4}/{\rm km}$     
    & $11.95_{-0.94}^{+1.04}$ & $11.89_{-0.93}^{+1.15}$
    & $11.28_{-0.62}^{+0.59}$ & $11.60_{-0.68}^{+0.81}$ 
    & $12.06_{-0.84}^{+0.89}$ & $12.51_{-0.99}^{+0.65}$
    \\ 
     $\mathcal{B}_{\rm NS}^{\rm HS}$     
    &1.78 &2.80 & $4.48\times10^4$ & 95.20& 3.72 &1.40 \\ 
        \hline
  \end{tabular*}
  \label{tb:post_prop}
\end{table*}

\begin{figure*}
        \includegraphics[width=7in]{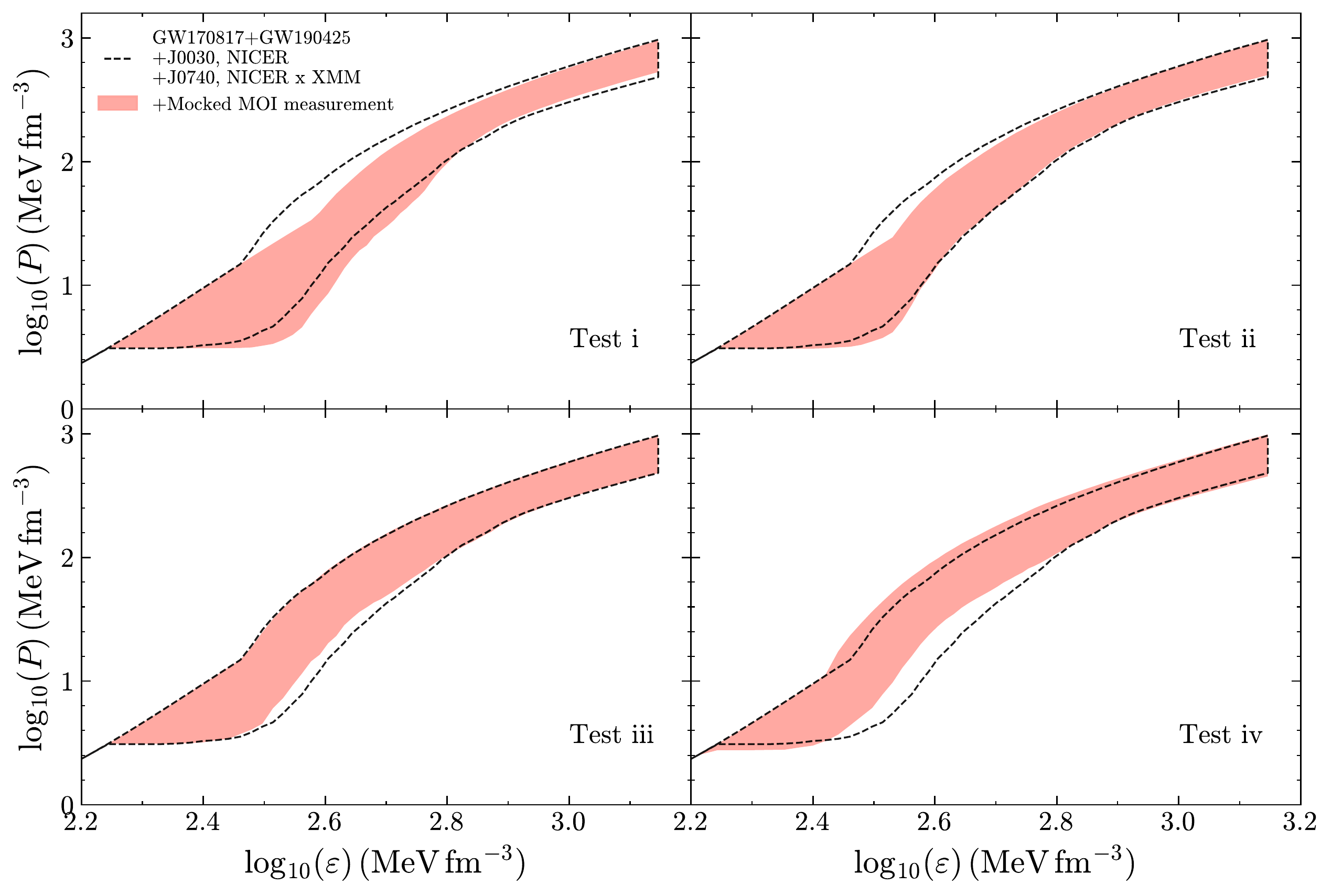}
         \vskip-2mm
    \caption{The 90\% credible regions of the EOSs for QMF+CSS. The dashed lines show the results from joint analysis of the present data, including two gravitational wave data and two \nicer's mass-radius measurements. The red shaded regions show the updated posterior distributions from four tests (i-iv), in which we add a mocked MOI measurement for PSR J0737-3039 A to be $I/M_{\rm A}^{3/2}=35, 40, 45, 50\,{\rm km^2}/\Msun^{1/2}$, respectively.
    }
    \vspace{-0.4cm}
    \label{fig:eos}
\end{figure*}

\begin{figure*}
        \includegraphics[width=7in]{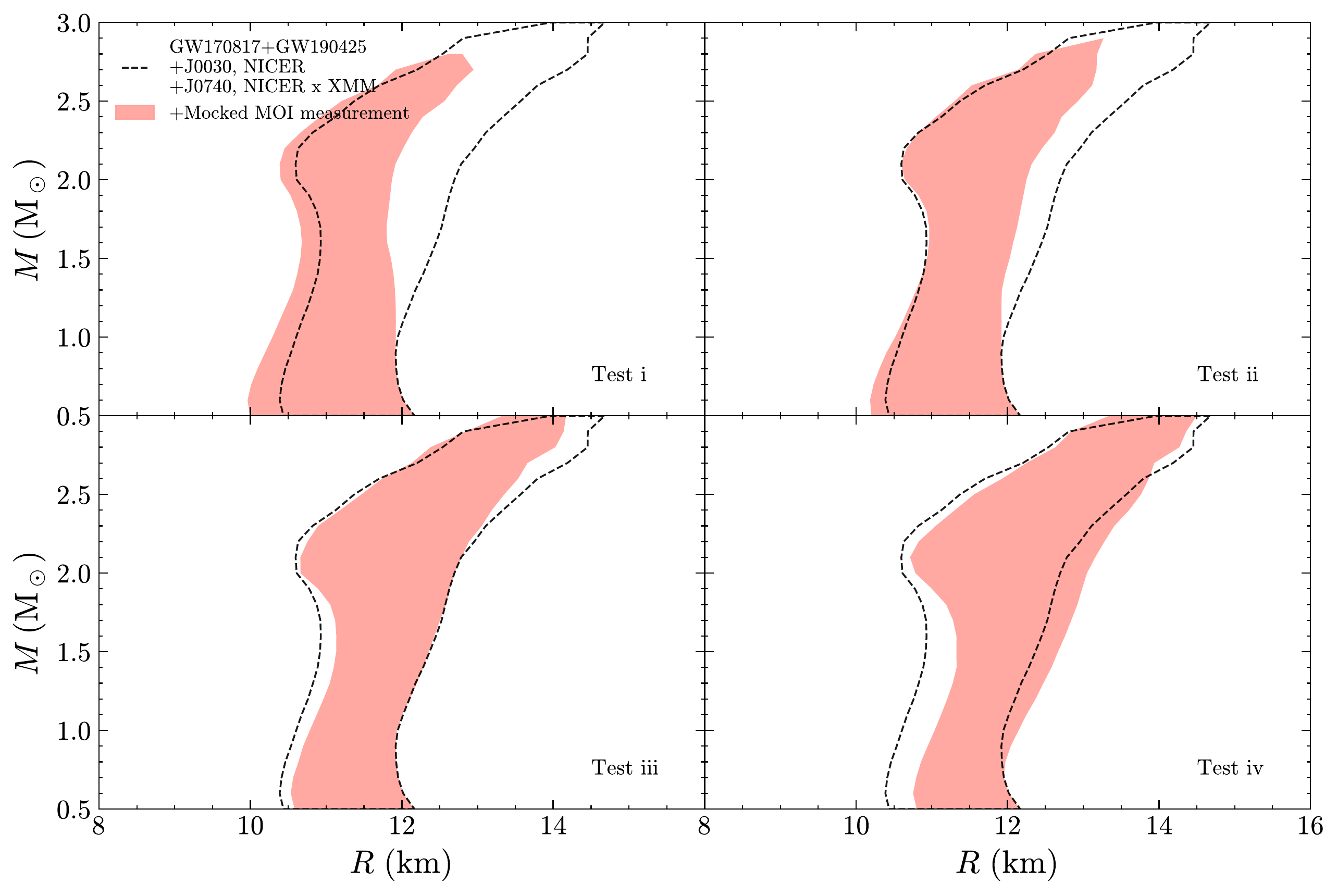}
         \vskip-2mm
    \caption{Same as Fig.\ref{fig:eos}, but for the mass-radius relations.
    }
         \vspace{-0.4cm}
    \label{fig:M-R}
\end{figure*}

\subsection{Updated constraints with mocked MOI measurements of PSR J0737-3039 A}
\label{sec:artI}

The MOI of A is expected to be measured with 11 percent precision at a 68\% confidence level in 2030~\citep{2020MNRAS.497.3118H}, and the results will provide further constraints on the EOS and neutron star properties.
To determine how such a precise MOI measurement would improve our present knowledge of the mass-radius relation, we carry out four tests (Tests i-iv) incorporating mocked data for future measurements of the MOI for A into our Bayesian analysis framework.

For Test i to Test iv, we add MOI measurements scaled by $M_{\rm A}^{3/2}$ as $I/M_{\rm A}^{3/2}=35, 40, 45, 50\,{\rm km^2}/\Msun^{1/2}$ (or equivalently, $I = 1.08, 1.23, 1.39, 1.54\times10^{45}\,{\rm g\ cm^2}$), with a 10\% accuracy for all four cases. 
We repeat the sampling process for the neutron stars and report the results in Table~\ref{tb:post_prop}, that is, the sound speed squared in quark matter $c_{\rm QM}^2$, 
the maximum mass $M_{\rm TOV}$ and the radius of a typical $1.4\Msun$ star $R_{1.4}$ for hybrid stars, along with results from~\citet{2021ApJ...913...27L}.
The 90\% credible regions of the EOSs are shown in Fig.~\ref{fig:eos} with and without the mocked MOI measurement for the representative QMF+CSS case. Fig.~\ref{fig:M-R} shows the corresponding results for the mass-radius relations.

We found in~\citet{2021ApJ...913...27L} that the maximum mass of neutron stars is $M_{\rm TOV}= 2.36_{-0.26}^{+0.49}\Msun$ ($2.39_{-0.28}^{+0.42}\Msun$) and the radius of a $1.4\Msun$ star is $R_{1.4} = 11.70_{-0.74}^{+0.85}\,{\rm km}$ ($11.95_{-0.94}^{+1.04}\,{\rm km}$) at 90\% credible interval for QMF+CSS (DD2+CSS) by incorporating both the GW170817 data and the \nicer~ mass-radius measurement of PSR J0030+0451, as well as the $2.14\Msun$ mass measurement of PSR J0740+6620 from radio timing.
In the present study, the peak value of the maximum mass shifts from $\sim 2.4\Msun$ to $\sim 2.2\Msun$ by updating the PSR J0740+6620 data with a more restrictive mass-radius measurement from the \nicer~x XMM data~\footnote{Apart from updating the PSR J0740+6620 data, there are two other differences between this study and \citet{2021ApJ...913...27L}: 1) the likelihood of encapsulating gravitational wave data obtained using the interpolation function and 2) additional GW190425 data. For the former, we checked that the results obtained using the interpolation function are consistent with those obtained using gravitational wave data, whereas for the latter, we proved that the addition of GW190425 data had a small effect on the results. Therefore, we consider the main difference between this study and \citet{2021ApJ...913...27L} to be the PSR J0740+6620 data update.}, indicating that a softer EOS is required.
This result can also be directly inferred from the posterior distribution of the sound speed squared in quark matter $c_{\rm QM}^2$, whose median shifts from $\sim 0.8$ to $\sim 0.7$.
See e.g., \citet{2021ChPhC..45e5104X} for more discussions on the sound speed.
As the resulting maximum mass is insensitive to the hadronic EOS~\citep{2021ApJ...913...27L}, we
use the two posterior sample sets (QMF+CSS and DD2+CSS) to re-generate two equal size sample sets and combined them as the weighted-average posterior samples.
We finally find $M_{\rm TOV} = 2.23_{-0.21}^{+0.41}\Msun$ at 90\% credible interval.
Such results suggest that the O3b compact object GW200210\_092254~\citep{2021arXiv211103606T}, with a mass around $2.8\Msun$, should be a black hole rather than a neutron star.
There is also a slight decrease in the radius due to this update. The inferred radius changes from $11.70_{-0.74}^{+0.85}\,{\rm km}$ ($11.95_{-0.94}^{+1.04}\,{\rm km}$) to $11.61_{-0.76}^{+0.75}\,{\rm km}$ ($11.89_{-0.93}^{+1.15}\,{\rm km}$) for QMF+CSS (DD2+CSS) at 90\% credible interval.

Next, we determine the effects of incorporating a mocked MOI measurement into the data.
Fig.~\ref{fig:eos} shows that the additional MOI measurement provides further constraints on the EOS.
These constraints on the EOS mainly occur at intermediate densities $\sim300-600\,{\rm MeV/fm^{3}}$ (or equivalently, $\sim2-4~\varepsilon_0$, where $\varepsilon_0\approx157\,{\rm MeV/fm^{3}}$ is the nuclear saturation mass density).
Previously, \citet{2020ApJ...888...12M} varied the uncertainty of the mocked MOI measurements (from 5\% to 50\%) but fixed the central value to the currently most favoured value of MOI, and found a 10\% precision MOI measurement could further improve the estimation on the radii for neutron stars with masses of $M=1.8-2\Msun$ but had a negligible effect on the corresponding estimates for masses of approximately $M=1\Msun$. 
Their result is consistent with our present result for Test ii (shown in the upper right panel of Fig.~\ref{fig:M-R}), since the mocked MOI in the Test ii case is close to the most probable value obtained from the present analysis, $I/M_{\rm A}^{3/2}\sim 41~{\rm km}^2/\Msun^{1/2}$ (as reported earlier in Sec.~\ref{sec:MOI prediction}).
Presently, we varied the central value of the MOI measurement but fixed the uncertainty to 10\%.
The results of the other three tests based on different MOI measurements show that it is possible to improve our estimates of neutron star radii over the entire mass span.

Additionally, as detailed in Table \ref{tb:post_prop}, as the measured MOI increases, the inferred sound speed in quark matter decreases.
Taking the QMF+CSS case as an example, $c_{\rm QM}^2$ drops from $0.77_{-0.27}^{+0.21}$ at $I/M_{\rm A}^{3/2}=35\,{\rm km^2/\Msun^{1/2}}$ (Test i) to $0.70_{-0.30}^{+0.27}$ for a measured $I/M_{\rm A}^{3/2}=50\,{\rm km^2}/\Msun^{1/2}$ (Test iv) because  larger stellar radii have softer quark matter EOS (i.e., smaller sound speeds).
Finally, it is important to note that the MOI measurement has a limited effect on the resulting maximum mass.
The inferred maximum mass is found to be insensitive to the MOI measurement, with the most likely value of the maximum mass remaining at $\sim2.2\Msun$.

\subsection{With or without deconfined quarks?}
\label{sec:withq}
In this section, we analyze whether the current data favour a strong first-order phase transition inside neutron star inner cores.
For this purpose, we compute the Bayes factor between the EOS models with and without a phase transition, $\mathcal{B}_{\rm NS}^{\rm HS}$. 
The results are reported in the ninth and sixteenth rows of Table~\ref{tb:post_prop}.

In general, the Bayes evidences are comparable for both soft QMF and stiff DD2 cases, indicating that the current data are compatible with the possibility of a phase transition, namely the compact objects are likely hybrid stars, besides the pure hadronic nature as commonly assumed.
The results are nevertheless slightly dependent on the two hadronic EOSs employed.
Following the quantifying of the significance in \citet{kass1995}, when only considering available observational data, the possibility of a phase transition is possible for the stiff DD2 case (with $x\equiv2\ln(\mathcal{B}_{\rm NS}^{\rm HS})=2.05$), but not supported in the soft QMF case (with $x<0$).
The evidence of a phase transition for the DD2 case is strengthened if we include the MOI measurement in the data: Table~\ref{tb:post_prop} shows the large Bayes factors, especially for relatively low MOI measurements with $x>2$ for Test i, ii and iii.
Note that the recent PREX-II measurement of the neutron skin of $^{208}$Pb indicates a rather stiff symmetry energy~\citep{2021PhRvL.126q2503R}, resulting in a large predicted radius for a $1.4\Msun$ star, which may conflict with the tidal deformability constraint from GW170817~\citep{2018PhRvL.121p1101A}.
An early phase transition, e.g., at $\sim2n_0$~\citep{2021ApJ...913...27L}, could resolve this conflict, because the transition could lead to a lower radius from the softening of the EOS at high densities.
Supporting evidence for a phase transition also justifies our introduction of a quark core into the present analysis.

\subsection{(Hybrid) neutron stars or quark stars?}
\label{sec:q}

There is a theoretical possibility that three-flavor quark matter may be even strongly bound with respect to the $^{56}\rm Fe$, namely the energy per baryon $(E/A)_{\rm uds}$ is smaller than $930\mev$, as proposed by~\citep{1971PhRvD...4.1601B,Terazawa1979,1984PhRvD..30..272W}. \footnote{See \citet{2022arXiv220304798Y} for recent discussions on the stability of both strange and  non-strange quark matter.}
Since it has so far proved very difficult to unambiguously distinguish quark stars from normal neutron stars from astrophysical observations~\citep{2007PhR...442..109L}, all pulsar-like objects should be quark stars, if not all of them.
Following similar procedure for the (hybrid)neutron star analysis in Sec. \ref{sec:analysis}, we consider different structural model for A in the present section and redo the bayesian analysis from quark matter EOSs based on the available gravitational-wave and x-ray observational data. 
Also, the crustal effects on the mass-radius relations of quark stars are found to be at most 2-3\%~\citep{2021MNRAS.506.5916L}, we, therefore, consider the quark stars to be bare ones in the present analysis. Nevertheless, it is not clear at all whether quark stars have a crust or not. See discussion in \citet{2005PrPNP..54..193W}.

The quark star matter is composed of u, d, s quarks with the charge neutrality maintained by the inclusion of leptons.
Since the QCD equation of motion can not be solved on the lattice at finite baryon number density, we use perturbative QCD (which is applicable above $40n_0$) and an MIT-type bag model for the EOS of quark star matter, with the possible existence of superfluidity for quarks being included~\citep{2001PhRvD..64g4017A,2001PhRvL..86.3492R,2002PhRvD..66g4017L}.
Then three EOS parameters appear in the likelihood of Eq.~(\ref{eq:posterior}) in the quark star case as $\boldsymbol\theta = \{B_{\rm eff},a_4, \Delta\}$: the bag parameter $B_{\rm eff}$, the QCD correction coefficient $a_4$ and the quark pairing gap $\Delta$.
Following our previous studies \citep{2018PhRvD..97h3015Z,2021ApJ...917L..22M,2021MNRAS.506.5916L}, we assign wide boundaries to them as $B_{\rm eff}^{1/4}\in[125, 150]\,{\rm MeV}$, $\Delta \in[0, 100]\,{\rm MeV}$, $a_4\in[0.4, 1]$ as theoretically estimated.
We mention here that additional stability requirements should be imposed on the EOS parameters of quark stars: 
First, the energy per baryon for non-strange quark matter should satisfy $(E/A)_{\rm ud}\geq934\,\mev$ to guarantee the observed stability of atomic nuclei; Second, $(E/A)_{\rm uds}\leq930\,\mev$ is required, according to the hypothesis that strange quark matter is absolutely stable. 
The most probable values of the EOS parameters (with their $90\%$ confidence boundaries) are reported in Table \ref{tb:quark}, based on the joint analysis of two gravitational wave data and two \nicer's mass-radius measurements.

\begin{table}
  \centering
  \caption{Most probable intervals of the EOS parameters ($90\%$ confidence level) for quark stars. See text for details.
  }
    \setlength{\tabcolsep}{0.8pt}
\renewcommand\arraystretch{1.8}
\begin{tabular*}{\hsize}{@{}@{\extracolsep{\fill}}lc@{}}
\hline Parameters  &  GW170817+GW190425+PSR J0030+ PSR J0740\\
\hline $B_{\rm eff}^{1/4}/\rm MeV$ & $144.4_{-7.6}^{+4.6}$  \\
\hline $a_4$  & $0.73_{-0.21}^{+0.22}$  \\
\hline $\Delta/\rm MeV$   &$71.2_{-52.4}^{+24.5}$ \\ \hline 
\end{tabular*}
         \vspace{-0.2cm}
  \label{tb:quark}
\end{table}

\begin{figure}
        \includegraphics[width=3.3in]{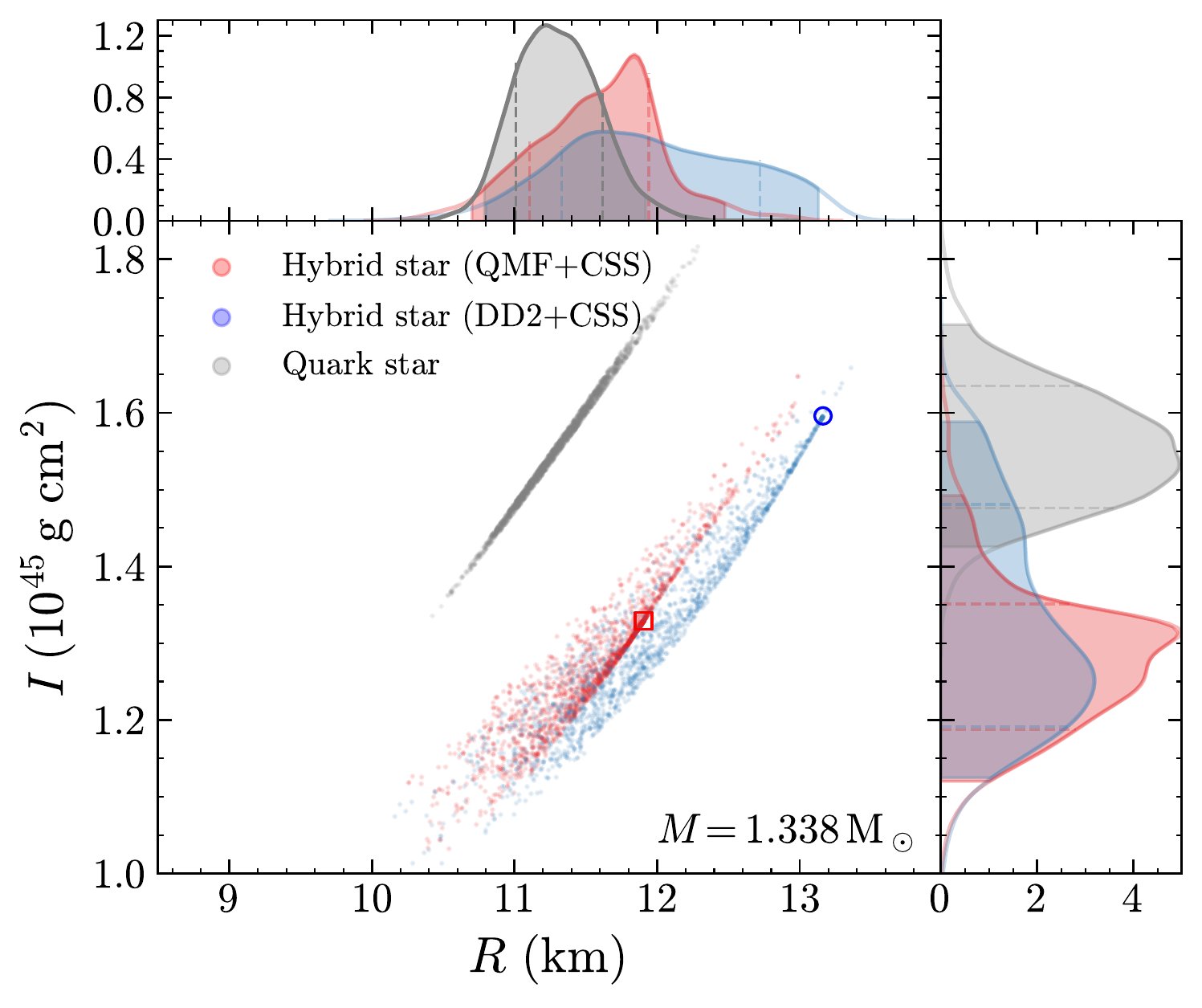}
    \caption{Posterior distributions of $I$ vs. $R$ for a $M=1.338\Msun$ star, transformed from the EOS parameters posteriors. The probability density of $R$ and $I$ are shown in the upper and right panel, respectively, with the 68\% credible interval in dashed lines and the 95\% credible interval in shade regions. Two markers denote the two results for pure neutron stars.
    }
             \vspace{-0.4cm}
    \label{fig:R-I_1338}
\end{figure}

\begin{table*}
  \centering
  \caption {Fit parameters of the $\bar I-C$ relation for hybrid stars and quark stars in the functional form of $\bar I = a_1C^{-1}+a_2C^{-2}+a_3C^{-3}+a_4C^{-4}$~\citep{2016MNRAS.459..646B}. The coefficients of determination are also listed in the last column.
  }
    \setlength{\tabcolsep}{0.8pt}
\renewcommand\arraystretch{1.3}
\begin{tabular*}{\hsize}{@{}@{\extracolsep{\fill}}lccccc@{}}
    \hline
     & $a_1$ & $a_2$ & $a_3$ &$a_4$ & $R^2$\\
    \hline
    Hybrid star & $6.759\times10^{-1}$&  $3.356\times10^{-1}$& $-1.820\times10^{-2}$ & $5.514\times10^{-4}$  &0.99757 \\ 
    Quark star & $3.461\times10^{-1}$ &$4.013\times10^{-1}$ &$-1.198\times10^{-3}$ & $6.499\times10^{-5}$ &0.99999\\    
    \hline
  \end{tabular*}
  \label{tb:fitcoeff}
\end{table*}

\begin{figure}
        \includegraphics[width=3.3in]{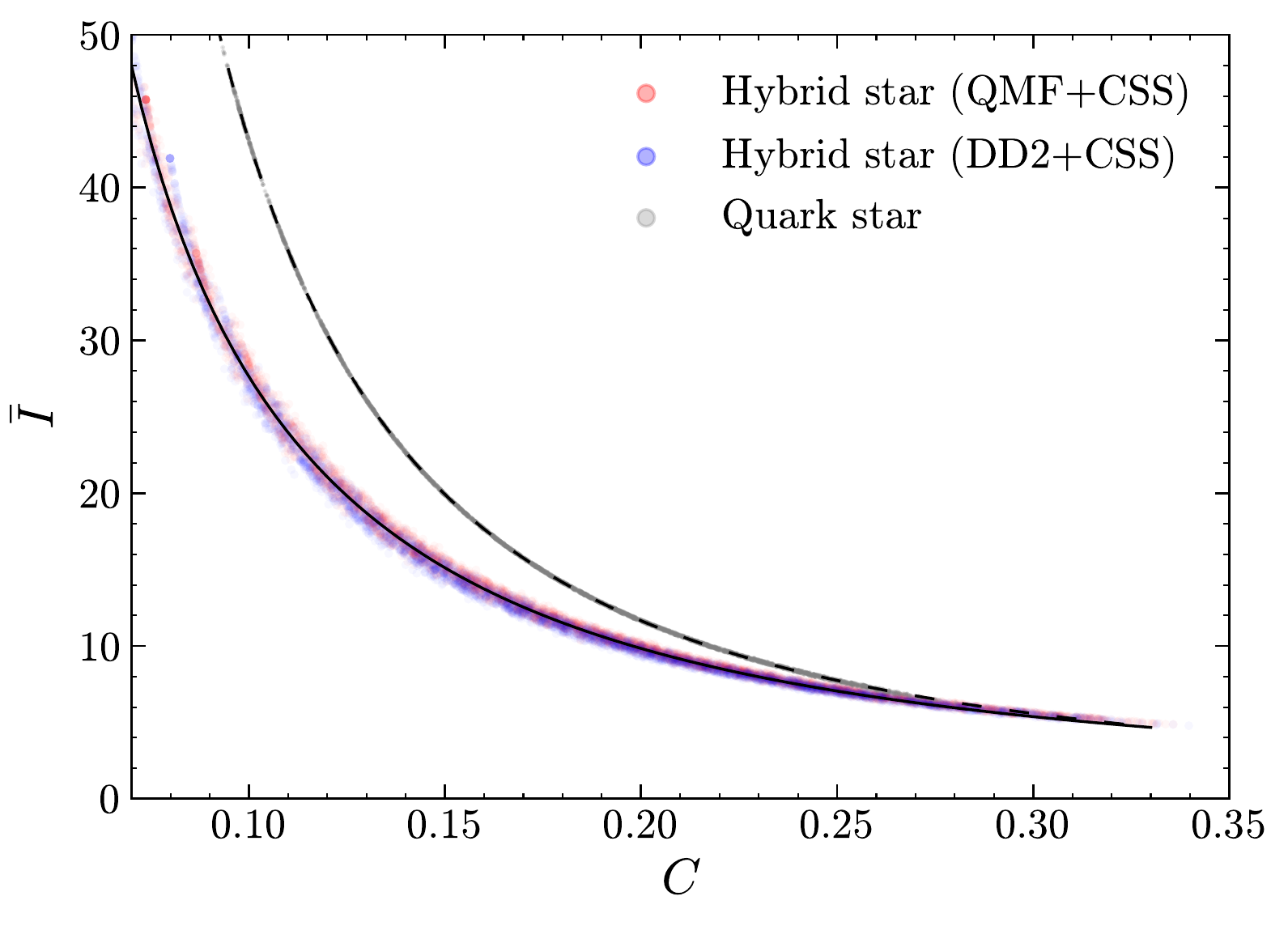}
    \caption{Posterior distribution of $\bar I \equiv I/M^3$ shown as a function of stellar compactness $C\equiv M/R$, transformed from the EOS parameters posteriors. The solid and dashed lines represent the fitting functions for the (hybrid-)neutron star (with both QMF+CSS and DD2+CSS cases) and quark star results, respectively. 
    }
         \vspace{-0.5cm}
    \label{fig:Ibar_compactness}
\end{figure}

In Fig.~\ref{fig:R-I_1338}, we show the posterior distributions of $I$ as functions of $R$ for both quark stars and (hybrid) neutron stars. 
One observes that the two $I$ vs.$R$ distributions are located in separate regions and become more and more separated with each other when increasing $R$ (equivalently decreasing the compactness $M/R$), suggesting that it can be possible to distinguish them from simultaneous measurement of the stellar radius and the MOI. 
In particular, we calculate the MOI of A as a quark star to be $I_A=1.55_{-0.11}^{+0.13}\times10^{45}\,{\rm g\,cm^2}$, at 90\% credible interval.
Such predicted results based on quark star EOSs are obviously larger than those based on neutrons star EOSs.
Consequently, If the MOI is measured large for A, for example, $I_A\gtrsim1.4\times10^{45}\,{\rm g\,cm^2}$, it is most likely a quark star rather than a neutron star with or without a quark core (see Table \ref{tb:MOIcomparison}), provided that the accuracy of the radius measurement is at least $\sim 1\,\km$.
Also, there is a good correlation between the radius and the MOI of quark stars, which can be fitted by ${I}/({10^{45}\,{\rm g\,cm^2}}) = aR^b$ with $a=1.69\times10^{-2}$, $b=1.87$ and the coefficient of determination is $R^2=0.997$. 
The fitted formula allows us to translate accurately from a MOI measurement to its radius and vice versa, under the condition of A being a quark star. 

Before the end of the section, we come back to the universal relations mentioned at the beginning of Sec.~\ref{sec:res}. 
It is expected that the dimensionless MOI $\bar{I}$ of quark stars follow the similar universal relations with neutron stars and hybrid stars when expressed as a function of $\Lambda$, as shown in, e.g.,~\citet{2017PhR...681....1Y}.
Nevertheless, the $\bar{I}$ values of quark stars should behave quite differently from those of neutron stars and hybrid stars when expressed as a function of the stellar compactness $C\equiv M/R$~\citep{1994ApJ...424..846R,2002A&A...396..917B,2005ApJ...629..979L}.
We show in Fig.~\ref{fig:Ibar_compactness} the dimensionless MOI $\bar{I}$ as a function of $C$ for both (hybird-)neutron stars and quark stars.
We see that the neutron star and quark star results are both well distributed in universal $\bar I$-versus-$C$ curves in the form of $\bar I = a_1C^{-1}+a_2C^{-2}+a_3C^{-3}+a_4C^{-4}$~\citep{2016MNRAS.459..646B} (the fit parameters are collected in Table~\ref{tb:fitcoeff}).
The two universal lines reconcile in the high-compactness regime, but it is evident that the (hybrid-)neuron star line deviates from the quark star one and moves downwards in the low-compactness regime.
This is because the matter in quark stars is relatively evenly distributed; The surface density is actually close to the central density, resulting in a larger MOI than in the (hybrid-)neutron star case when decreasing the stellar mass or compactness~\citep{2013PhRvD..88b3009Y}.

\section{Summary and conclusions}
\label{sec:summary}

There is a high expectation of an imminent precise measurement of the neutron star MOI for the very first time for the first double pulsar system PSR J0737-3039. 
Continuing efforts are being made to theoretically predict the MOI of PSR J0737-3039 A based on many modern nuclear EOSs within the constraints provided by the results of nuclear experiments and astrophysical observations.
Considering that the pulsar interior may have a quark matter core, it is meaningful to update the MOI analysis of previous studies by incorporating strangeness phase transitions.
Furthermore, such an MOI measurement would place new constraints on the neutron star EOS. A question that naturally follows is what information this measurement will contribute to our understanding of dense matter. 

In this study, a hadron-quark phase transition in the EOS prior modelled by the generic CSS parameterization is explicitly considered and a Bayesian analysis is performed on the MOI of PSR J0737-3039 A using available data from LIGO/Virgo and \nicer.
Both soft QMF and stiff DD2 are applied for the low-density hadronic phase to test the model-dependence of the MOI prediction. 
A MOI of $1.27_{-0.14}^{+0.18}\times10^{45}\,{\rm g\,cm^2}$ is obtained using QMF+CSS, and a similar result of $1.29_{-0.15}^{+0.26}\times10^{45}\,{\rm g\,cm^2}$ is obtained using DD2+CSS.
We also demonstrated in detail that an accurate MOI measurement, when available, can significantly reduce the EOS uncertainties, 
especially when the MOI value is  a few times the nuclear saturation density.
In addition, we evaluated the relative probability of a phase transition occurring using the pure hadronic star as a reference state, based on the available data and a potential MOI constraint.
The results show that the hybrid star scenario is justified as one of the many possibilities of the nature of pulsars, and the maximum mass is predicted to be $M_{\rm TOV}\sim2.2\Msun$. 
On the other hand, a relatively larger MOI result is found, $I_A\sim1.55\times10^{45}\,{\rm g\,cm^2}$, if assuming PSR J0737-3039 A is a quark star, compared to that of (hybrid-)neutron-star modelling, suggesting it is still possible to distinguish between the two if more precise radius measurement is possible (at least $\sim1\,\km$ accuracy), possible in the near future with joint efforts of the multimessenger missions, like aLIGO/Virgo, \nicer, eXTP, SKA. We finally examine the MOI–compactness universal relations and provide easy-to-use analytic fits on the results for both (hybrid-)neutron-star and quark star cases.

\section*{ACKNOWLEDGMENTS}
We are thankful to Prof. B.-A. Li and Tong Liu for helpful discussions. 
The work is supported by National SKA Program of China (No.~2020SKA0120300), the National Key Research and Development Program of China (No. 2017YFA0402600), the National Natural Science Foundation of China (Nos.~11873040 and 11833003), the science research grants from the China Manned Space Project (No. CMS-CSST-2021-B11) and the Youth Innovation Fund of Xiamen (No. 3502Z20206061).
\section*{Data Availability}

The data underlying this article are available in the article.

\end{document}